\documentclass[10pt,letterpaper,twocolumn]{article} %
\usepackage{ol2}                                    %
\usepackage{ol2}
\usepackage[draft]{hyperref}
\usepackage{amsmath,amssymb}

\begin{document}
\twocolumn[ %% activate for two-column option

\title{Four-wave mixing, quantum control and compensating losses in doped negative-index photonic metamaterials}

%% For REVTeX it is possible to automate superscript and e-mail callouts with the superscriptaddress option; see REVTeX4 documentation.

\author{Alexander K. Popov,$^{1,*}$ Sergey A. Myslivets,$^2$  Thomas F. George,$^{3}$ and Vladimir M. Shalaev$^4$ }

\address{
$^1$Department of Physics \& Astronomy, University of
Wisconsin-Stevens Point, Stevens Point, WI 54481, USA
\\
$^2$Institute of Physics of the Russian Academy of
Sciences, 660036 Krasnoyarsk, Russian Federation \\
$^3$Center for Nanoscience, Department of Chemistry \& Biochemistry
and Department of Physics \& Astronomy,\\
University of Missouri-St.~Louis, St.~Louis, MO 63121, USA\\
$^4$Birck Nanotechnology Center and School of Electrical and
Computer Engineering, \\Purdue
University,  West Lafayette, IN 47907, USA \\
$^*$Corresponding author: apopov@uwsp.edu}
\date{June 10, 2007}
\begin{abstract}
The possibility of compensating absorption in negative-index
metatamterials (NIMs) doped by resonant nonlinear-optical centers is
shown. The role of quantum interference and extraordinary properties
of four-wave parametric amplification of counter-propagating
electromagnetic waves in NIMs are discussed.
\end{abstract}

\ocis{1904410, 270.1670.}

] %% activate for two-column option

Negative refractive index metamaterials (NIMs) present a novel class
of materials that promise a revolutionary  breakthrough in
electromagnetics (for review, see, e.g.,  \cite{Sh}). Nonlinear
optics in such materials remains so far  a less developed branch of
optics. The possibility of nonlinear electromagnetic responses in
such materials attributed to the asymmetry of the voltage-current
characteristics of their building blocks was predicted in
\cite{Lap,Kiv}. Recent  experimental demonstrations of the exciting
opportunities to craft nonlinear optical materials with
characteristics exceeding those in natural crystals are reported in
\cite{Kl}. Unique nonlinear-optical (NLO) propagation effects
associated with three-wave ($\chi^{(2)}$) coupling in NIMs, as
compared with their well known  counterparts in natural materials,
were revealed in \cite{Agr,Sc,APB,OL}. The striking changes in the
optical bistability in a layered structure including a NIM layer
were shown in \cite{Lit}. A review of the corresponding theoretical
approaches is given in \cite{Gab}. The most detrimental obstacle
toward applications of NIMs is strong absorption that is inherent to
this class of materials. The possibility to overcome such obstacles
based on three-wave optical parametric amplification (OPA) in NIMs
was shown in \cite{APB,OL}. A great deal of technical problems must
be solved, however, in order to match the frequency domains of
negative index (NI),  strong NLO response and  the phase-matching to
realize such feasibility. Herewith, we propose and explore an
alternative approach associated with \emph{resonant four-wave mixing
(FWM) nonlinearities $\chi^{(3)}$ embedded in NIMs and tailored
through quantum control}. The possibility of compensating losses and
manipulating transparency, refractive index and nonlinear response
of the NIM sample with  control laser(s) is shown.

\begin{figure}[!h]
\begin{center}
\includegraphics[height=.45\columnwidth]{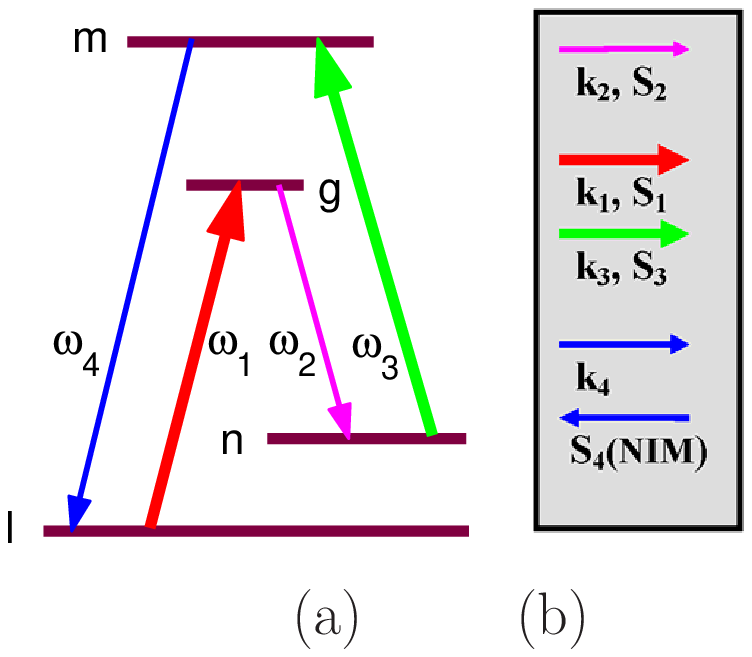}
%\includegraphics[height=.35\columnwidth]{Popov_f1aCol.eps}
%\includegraphics[height=.35\columnwidth]{Popov_f1bCol.eps}\\
%\hspace{12 mm}(a) \hspace{12 mm} (b)
\end{center}
\caption{Scheme of quantum-controlled FWM interaction (a) and
coupling geometry (b). $\omega_4$ is signal frequency, $\omega_2$ is
idler, and $\omega_1$ and $\omega_3$ are control fields.
$n(\omega_4)< 0$. \label{f1}}
\end{figure}

The basic idea of the proposed approach is as follows. A slab of NIM
is  doped by  four-level nonlinear centers [Fig. \ref{f1}(a)] so
that the frequency $\omega_4$ falls in the NI domain, whereas all
the other frequencies are in the  the positive index domain. Below,
we show the feasibility to produce the transparency and
amplification for the signal wave at $\omega_4$ controlled by two
lasers at $\omega_1$ and $\omega_3$. These three fields also
generate an idler at $\omega_2=\omega_3+\omega_1-\omega_4$, which
experiences either ordinary, population-inversion, or Raman
amplification provided by the driving field at $\omega_1$ and
controlled by another driving field at $\omega_3$. The amplified
idler contributes back to $\omega_4=\omega_3+\omega_1-\omega_2$
through FWM which leads to strongly enhanced OPA.
We assume that the wave vectors of all waves, $\mathbf{k}_j$, are
co-directed, which is required for phase-matching. Since only
$\omega_4$ experiences negative refraction and all other frequencies
are in the positive index  domain, energy flow at $\omega_4$ is
counter-directed against other waves. Hence, this signal wave must
enter the slab from its opposite side at $z=L$, which is the exit
facet for all other waves entering the slab at $z$=0
[Fig.\ref{f1}(b)]. Correspondingly, the signal exits the slab at
$z$=0. Such backwardness of the signal wave leads to
counterintuitive, distributed feedback-like behavior of the OPA
process in NIMs described in \cite{APB,OL}.

The equations describing the coupled waves at $\omega_4$ and
$\omega_2$ are as follows
\begin{eqnarray}
{da_{4}}/{dz} &=&-i\gamma_4^{(3)} a_{2}^{\ast }\exp [i\Delta
kz]+({\alpha_{4}}/{2}
)a_{4},  \label{a11} \\
{da_{2}}/{dz} &=&i\gamma_2^{(3)} a_{4}^{\ast }\exp [i\Delta
kz]-({\alpha_{2}}/{2 })a_{2}. \label{a21}
\end{eqnarray}
Here
$\gamma_{2,4}^{(3)}=(\sqrt{\omega_4\omega_2}/\sqrt[4]{\epsilon_4\epsilon_2/\mu_4\mu_2})
({4\pi}/{c})\chi_{2,4}^{(3)}E_{1}E_{3}$ are NLO coupling
coefficients; $\epsilon_j$ and $\mu_j$ are the dielectric
permittivities and magnetic permeabilities (which are negative at
$\omega_4$); $\Delta k=k_{1}+k_{3}-k_{2}-k_{4}$; and $\alpha_{j}$
are the absorption  or amplification coefficients. Photon fluxes are
given by $S_{2,4}/\hbar\omega_{2,4}=(c/8\pi)|a_{2,4}|^2$. The
amplitudes of the fundamental (control) waves $E_{1}$ and $E_{3}$
are assumed constant along the slab. Transmittance (amplification)
at $\omega_4$ is given by the factor $\eta_{4a}=|a_4(0)/a_4(L)|^2$,
where $L$ is the slab thickness. Note that the signs in (\ref{a11})
are opposite to those in ordinary media, which is due to the
backwardness of the signal wave. The solution to a similar set of
equations and its analysis are given in \cite{APB,OL}.

\begin{figure}[!h]
\begin{center}
\includegraphics[width=.8\columnwidth]{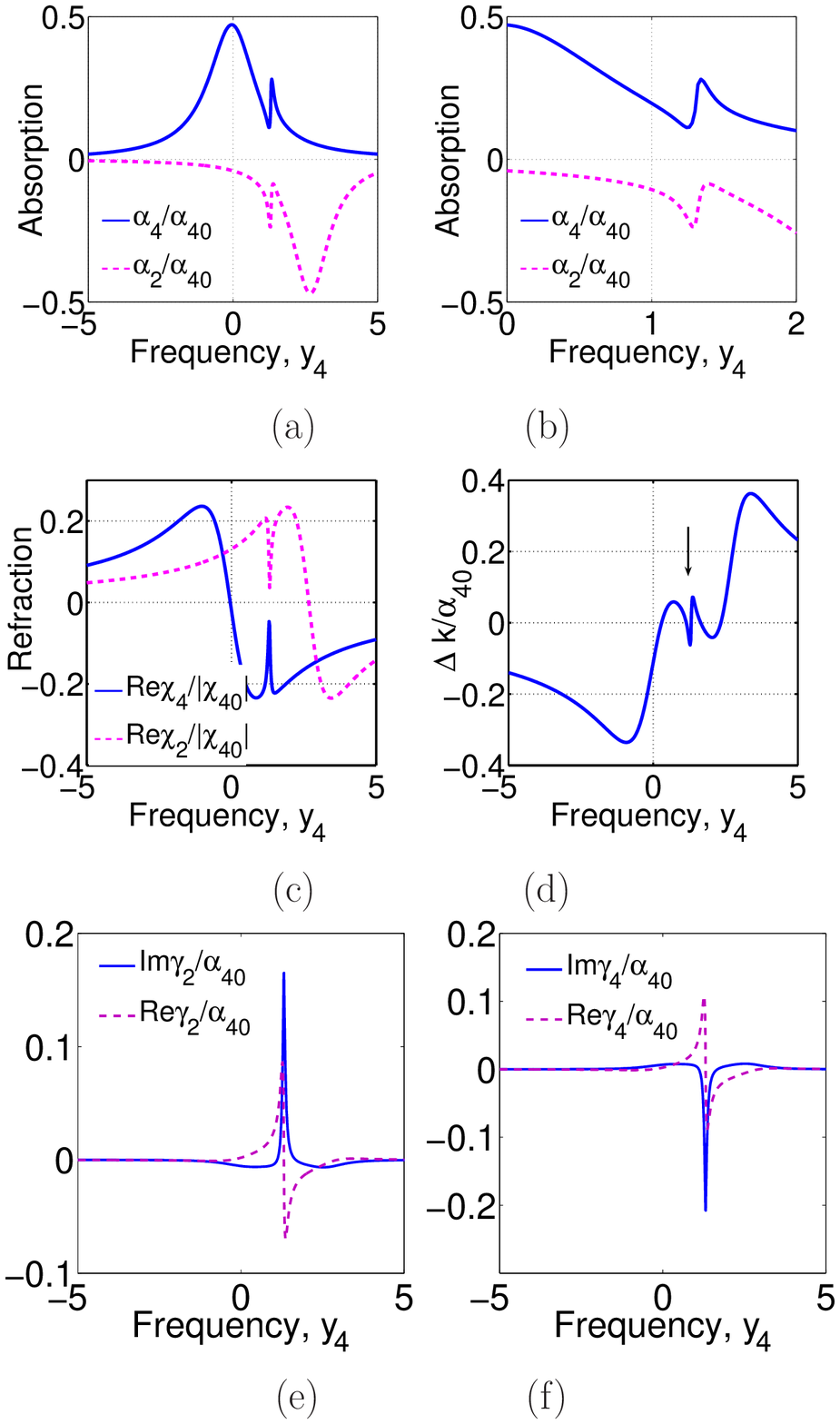}
%\includegraphics[width=.38\columnwidth]{Popov_f2a.eps}
%\includegraphics[width=.38\columnwidth]{Popov_f2b.eps}\\
%(a) \hspace{28mm} (b)\\
%\includegraphics[width=.38\columnwidth]{Popov_f2c.eps}
%\includegraphics[width=.38\columnwidth]{Popov_f2d.eps}\\
%(c) \hspace{28mm} (d)\\
%\includegraphics[width=.38\columnwidth]{Popov_f2e.eps}
%\includegraphics[width=.38\columnwidth]{Popov_f2f.eps}\\
%(e) \hspace{28mm} (f)
\end{center}
\caption{Nonlinear interference resonances induced by the control
fields with $G_1=G_3=50$~GHz and  $\Omega_1=\Omega_3=2.5
\Gamma_{lg}$. ($y_4 = \Omega_4/\Gamma_{ml}$).\label{chi}}
\end{figure}

Calculations of the optical constants for embedded NLO centers
(driven by the control fields) with account for constructive and
destructive quantum interference are performed following the
density-matrix technique described in \cite{PPRA}. In our
simulations, we used the following representative values for
relaxation rates: energy level relaxation rates $\Gamma_n=20 \times
10^6$~s$^{-1}$, $\Gamma_g=\Gamma_m=120 \times 10^6$~s$^{-1}$;
partial transition probabilities $\gamma_{gl}=7\times
10^6$~s$^{-1}$, $\gamma_{gn}=4\times 10^6$~s$^{-1}$,
$\gamma_{mn}=5\times 10^6$~s$^{-1}$, $\gamma_{ml}=10\times
10^6$~s$^{-1}$; homogeneous transition half-widths
$\Gamma_{lg}=10^{12}$~s$^{-1}$,  $\Gamma_{lm}=1.9 \times
10^{12}$~s$^{-1}$, $\Gamma_{ng}=1.5 \times 10^{12}$~s$^{-1}$,
$\Gamma_{nm}=1.8 \times 10^{12}$~s$^{-1}$, $\Gamma_{gm}=5 \times
10^{10}$~s$^{-1}$; $\Gamma_{ln}=10^{10}$~s$^{-1}$. We assumed that
$\lambda_2=756$ nm, $\lambda_4=480$ nm. The results of numerical
simulations for the optical coefficients entering equations
(\ref{a11}) and (\ref{a21}) are shown in Fig. \ref{chi}. Here,
$\Omega_4=\omega_4-\omega_{ml}$; other resonance detunings
$\Omega_{j}$ are defined in a similar way. The coupling Rabi
frequencies are introduced as $G_{1}= E_1d_{lg}/2\hbar$ and $G_{3} =
E_3d_{nm}/2\hbar$.  The quantities $\chi_{4,2}$ are effective linear
susceptibilities and $\alpha_{40}$ and $\chi_{40}$ denote their
resonant values when all the driving fields are turned off.
Figures~\ref{chi}~(a,b) depict changes in the absorption and
amplification coefficients for the signal  and idler, respectively,
produced by the control fields. (Fig.~\ref{chi}~(b) shows the
expanded interval corresponding to nonlinear interference
resonances.) Figure~\ref{chi}~(c) displays changes in refractive
indices, and Fig.~\ref{chi}~(d) shows wave-vector mismatch. The
arrow in Fig.~ \ref{chi}~(d) points out the interval of $y_4$,
roughly between 0.29 and 2.34, with five detunings $y_4$ for which
$\Delta k =0$ (under the assumption that the host material does not
introduce any additional phase mismatch). Figures~\ref{chi}~(e) and
\ref{chi}~(f) show narrow resonances in NLO coupling coefficients,
$\gamma_{4,2}^{(3)}$, with the width on the order of other
interference resonances. Figure~\ref{chi} proves the feasibility of
manipulating local optical parameters through nonlinear quantum
interference induced by the control fields. For the used optical
transitions rates, the magnitude of $G \sim 10^{12}$~s$^{-1}$
corresponds to control field intensities $I$ of 10 to 100
kW/(0.1mm)$^2$.
\begin{figure}[!h]
\begin{center}
\includegraphics[width=.8\columnwidth]{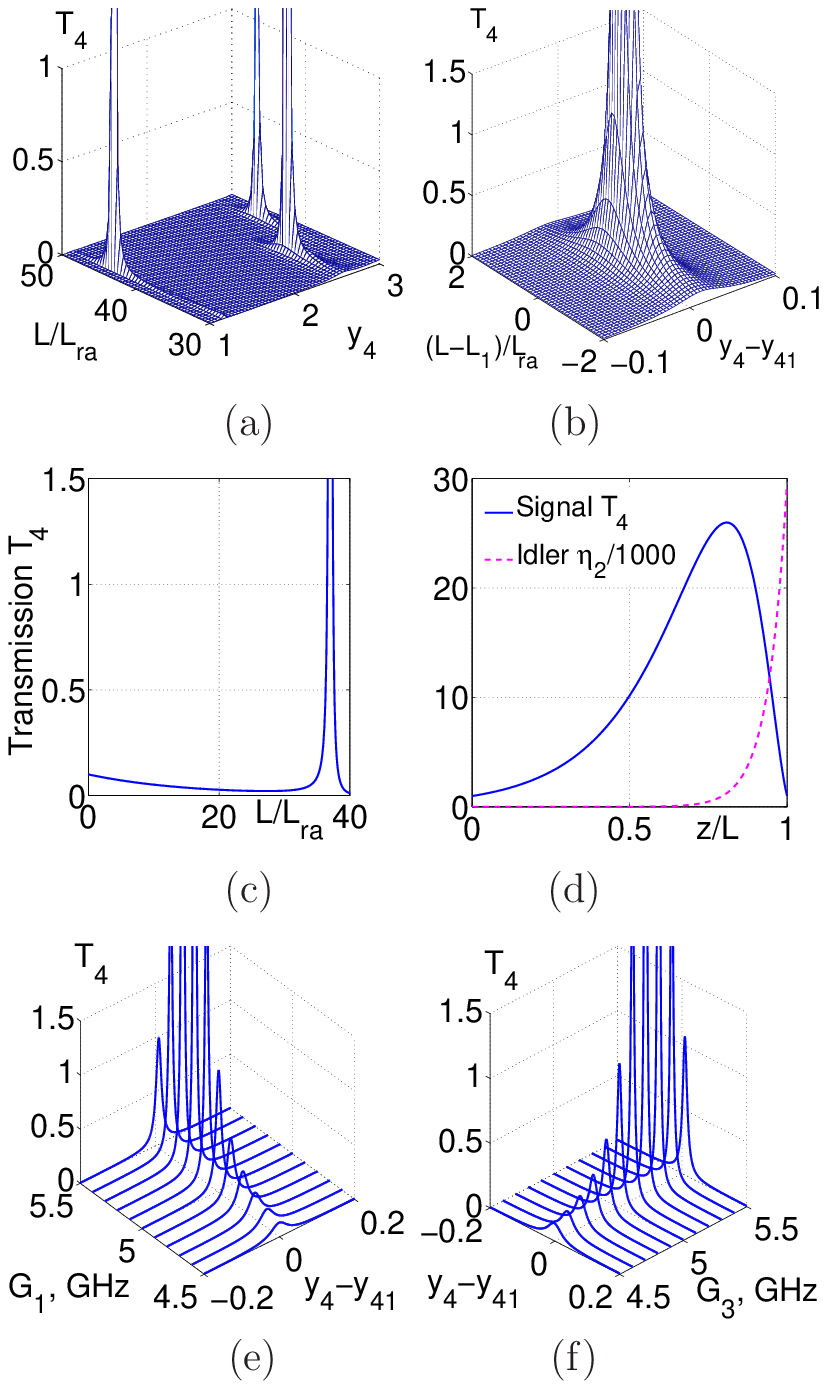}
%\includegraphics[width=.2\textwidth]{Popov_f3a.eps}
%\includegraphics[width=.2\textwidth]{Popov_f3b.eps}\\
%(a) \hspace{25mm} (b)\\
%\includegraphics[width=.2\textwidth]{Popov_f3c.eps}
%\includegraphics[width=.2\textwidth]{Popov_f3d.eps}\\
%(c) \hspace{25mm} (d)\\
%\includegraphics[width=.2\textwidth]{Popov_f3e.eps}
%\includegraphics[width=.2\textwidth]{Popov_f3f.eps}\\
%(e) \hspace{25mm} (f)\\
\end{center}
\caption{Laser-induced transmission resonances in the NI frequency
domain. $\Omega_1=\Omega_3=2.5\cdot\Gamma_{lg}$. (a)-(d):
$G_1=G_3=50$~GHz. (a)-(c), (e)-(f): $z = 0$. (c)-(d):
$y_4=y_{41}\equiv 2.5266$. (c): maximum is at
$\alpha_{40}L=L/L_{ra}=L_1/L_{ra}\equiv37.02$, $T_4\approx1$ at
$L/L_{ra}=36.52$. (d): intensity distribution inside the slab:
signal -- solid line, idler -- dashed line; $L/L_{ra}=36.52$. (e)
and (f): $L = L_1$. (e): $G_3 = 50$ GHz. (f): $G_1 = 50$ GHz.
\label{T}}
\end{figure}

The optimization of the output signal at $z=0$ is determined by the
interplay between absorption, idler gain and FWM, with the later
depending on the wave vector mismatch. This is a multi-parameter
problem involving sharp resonance dependencies. Results of our
numerical analysis of the steady-state solutions \cite{PPRA} to the
density matrix equations and Maxwell's equations for the
slowly-varying amplitudes in (\ref{a11}) and (\ref{a21}) are shown
in Fig.~\ref{T}. The transmission of the host slab in the NI
frequency domain has been  set as 10\%. Unlike conventional media,
the output signal for the waves coupled in the NIM slab through OPA
represents a set of distributed feedback-type resonances
\cite{APB,OL}. Such resonant behavior can be observed as a function
of the intensity of the fundamental fields,  the product of the slab
length and the density of NLO centers, and the resonance offsets for
the signal and fundamental fields. Figure \ref{T}(a) shows these
narrow transmission resonances. Here, we introduce the scaled
product of the slab length and the  density number of embedded
centers, $L/L_{ra}$, through the resonance absorption length,
$L_{ra}=\alpha_{40}^{-1}$. Figure~\ref{T}(b) displays the second
peak in Fig.~\ref{T}(a) with greater detail. It shows that the
transparency window is on the scale of the narrowest (Raman, in this
case) transition half-width, as a function of detuning, and the
resonant absorption length, as a function of length.
Figure~\ref{T}(c) depicts the same dependence at the optimum
resonance offset. Figure \ref{T}(d) shows that the intensity of the
signal inside the slab may significantly exceed its output value at
$z$=0, which depends on the ratio of the OPA and absorption rates.
Here, $\eta_2(z)=\left\vert {a_{2}(z)}/{a_{4L}}\right\vert ^{2}$.
Remarkably, the transparency and amplification occur in the
frequency range where magnitudes of $|\gamma_{4,2}|$ are
substantially less than their  resonant values. Figures \ref{T}(e)
and \ref{T}(f) display similar resonance dependence on the strength
of the control fields. The amplification in the maximums in
Fig.~\ref{T} reaches many orders of magnitude, which indicates
\emph{the feasibility of oscillations} and, hence, the generation of
counter-propagating left-handed signal and right-handed idler
photons. It is known that even small amplification per unit length
may lead to lasing provided that there is  a high-quality cavity (or
feedback resonance) which effectively increases the distance over
which the amplification occurs.

According to Fig.~\ref{T}(c), characteristic values of
$\alpha_{40}L~\sim~10$ are required to ensure the transparency and
gain. Assuming $\sigma_{40}~\sim~10^{-16}$~cm$^2$ for the resonance
absorption cross-section, which is typical for dye molecules, and
N~$\sim~10^{19}$ cm$^{-3}$ for the density of molecules, we obtain
that $\alpha_{40}~\sim 10^3$ to $10^4$~cm$^{-1}$, and the required
slab thickness in the range of L~$\sim$10 to 100$~\mu m$. At these
values, the contribution of the nonlinear centers in the refraction
index is estimated as $\Delta n < 0.5(\lambda/4\pi)\alpha_{40}\sim
10^{-2}-10^{-3}$, which essentially does not change the linear
negative refractive index.

In conclusion, we propose the compensation of losses in strongly
absorbing NIMs through  embedded tailored optical nonlinearities.
Such a possibility is shown with a realistic numerical model. We
have studied the resonant FWM-based OPA in such composite
metamaterials with a negative refractive index at the frequency of
the signal and a positive index for all other  coupled waves. The
strong nonlinear optical response of the composite is primarily
determined by the embedded  four-level nonlinear centers and, hence,
can be adjusted independently. In addition, we have shown the
possibility of quantum control of the local optical parameters,
which employs constructive and destructive quantum interference
tailored by two auxiliary control fields. Frequency-tunable
transparency windows in the negative-index frequency domain,
cavity-free generation of entangled counter-propagating photons, and
the feasibility of quantum switching in NIMs have been shown.

This work was supported in part by  ARO award W911NF-07-1-0261 and
ARO-MURI award 50342-PH-MUR.


\begin{thebibliography}{99}

\bibitem{Sh} V. M. Shalaev, Nature Photonics \textbf{1},
41-48 (2007).

\bibitem{Lap} M. Lapine, M. Gorkunov and K. H. Ringhofer,
Phys. Rev. E \textbf{67}, 065601(1-4) (2003).

\bibitem{Kiv} A. A. Zharov, I. V. Shadrivov, and Y. S. Kivshar, Phys. Rev. Lett.
\textbf{91}, 037401(1-4) (2003).

\bibitem{Kl} M. W. Klein, M. Wegener, N. Feth and S. Linden, Opt. Express \textbf{15}, 5238-5247 (2007).

\bibitem{Agr} V. M. Agranovich, Y.R. Shen, R.H. Baughman and A. A. Zakhidov,
Phys. Rev. B \textbf{69}, 165112(1-7) (2004).

\bibitem{Sc}  M. Scalora, G. D'Aguanno, M. Bloemer,  M. Centini,
N. Mattiucci, D. de Ceglia, and  Yu. S. Kivshar, Opt. Express
\textbf{14}, 4746-4756 (2006).

\bibitem{APB} A. K. Popov and V. M. Shalaev, Appl. Phys. B \textbf{84}, 131-137 (2006).

\bibitem{OL} A. K. Popov and V. M. Shalaev, Opt. Lett. \textbf{31}, 2169-2171 (2006).

\bibitem{Lit} N. M. Litchinitser, I. R. Gabitov, A. I. Maimistov, and V. M.
Shalaev, Opt. Lett. \textbf{32},  151-153 (2007)


\bibitem{Gab} A. I. Maimistov and I. R. Gabitov, http://arxiv.org/abs/nlin/0702023.

\bibitem{PPRA}  A. K. Popov, S. A. Myslivets, and T. F. George, Phys. Rev. A \textbf{71},
043811(1-13) (2005).

\end{thebibliography}
\end{document}